\documentclass[prb,overcite,preprint,endfloats]{revtex4-1}
\usepackage{graphicx}% Include figure files
\usepackage{dcolumn}% Align table columns on decimal point
\usepackage{bm}% bold math
\usepackage{color}

\begin{document}

%\preprint{Draft}

\title{Pd magnetism induced by indirect interlayer exchange coupling}

\author{W.E. Bailey}\footnote{web54@columbia.edu}
\affiliation{Materials Science \& Engineering, Dept. of Applied Physics \& Applied Mathematics, Columbia University, New York NY 10027, USA}
\author{ A. Ghosh, S. Auffret, E. Gauthier, U. Ebels}
\affiliation{SPINTEC, UMR(8191) CEA / CNRS / UJF / Grenoble INP ; INAC, 17 rue des Martyrs, 38054 Grenoble Cedex, France}
\author{ F. Wilhelm, A. Rogalev}
\affiliation{European Synchrotron Radiation Facility (ESRF), 38054 Grenoble Cedex, France}

\date{\today}
\begin{abstract}
We show that very large paramagnetic moments are created in ultrathin Pd layers through indirect interlayer exchange coupling.  Pd $L$-edge x-ray magnetic circular dichroism measurements show Pd moments in [Pd(2.5nm)/Cu(3nm)/Ni$_{81}$Fe$_{19}$(5nm)/Cu(3nm)]$_{20}$ superlattices which are ferromagnetically aligned with the applied field and nearly 3\% the size of Pd moments created in directly exchange coupled [Pd(2.5nm)/Ni$_{81}$Fe$_{19}$(5nm)]$_{20}$ superlattices.  The induced moment is two orders of magnitude larger than that expected from RKKY exchange acting on the bulk paramagnetic susceptibility of Pd. \end{abstract}

%\pacs{Valid PACS appear here}
%\keywords{Suggested keywords}
\maketitle

Pd is the elemental paramagnet closest to ferromagnetic order.  Its paramagnetic susceptibility $\chi$ and susceptibility enhancement factor $F$ over the Pauli band susceptiblity $\chi_0$, $F\equiv \chi/\chi_0\simeq\textrm{9.2}$\cite{l-b-4d5d}, is higher than any other $4d$ or $5d$ metal at low temperature, attributed to ferromagnetic correlations\cite{stonerPRS36,shimizuRPP81}.  Ferromagnetic correlations have also been invoked to explain the electronic specific heat\cite{schindlerPR67} and absence of superconductivity\cite{berkPRL66} in Pd.  Ferromagnetic ordering in ultrathin Pd films has long been predicted by density functional theory\cite{hongJAP88}, and giant susceptibilities have been observed in ultrathin Au/Pd/Au films\cite{brodskyPRL80}.

Direct exchange interactions from $3d$ ferromagnetic moments are known to induce sizeable local moments on Pd atoms.  Neutron diffraction\cite{cableJAP62} shows up to 0.4 $\mu_B$/Pd atom in Pd$_{3}$Fe.  Pd moments of similar magnitude are also induced by direct exchange at interfaces with Fe in ultrathin Pd/Fe superlattices.  Here, control over the Fe-Fe spacing allows determination of the moment profile in Pd.  $L$-edge x-ray magnetic circular dichroism (XMCD) measurements\cite{tholePRL1992} show 0.4 $\mu_B$/interface atom\cite{vogelPRB97}, decaying to half that value by the third monolayer.

Indirect exchange interactions have not yet been shown to have an effect on the magnetism of Pd.  Pd is among very few $3d$, $4d$, or $5d$ non-ferromagnetic transition metals (N) through which no oscillatory interlayer exchange coupling (IEC, or RKKY-like\cite{rudermanPR54}) between adjacent ferromagnets (F)\cite{parkincoupling} has been observed in a F/N/F structure; Pt is another example\cite{liPRB94}.  The tendency towards ferromagnetic order in Pd might explain the absence of antiferromagnetic coupling in F/Pd/F multilayers, but it might also seem to predict the possibility of ferromagnetic order induced through a F/N/Pd structure.  However, an all-electronic (or -magnetic) search for such coupling would seem to be precluded by the lack of magnetic moment in Pd.

Static exchange coupling in F/Cu/N$_2$=(Pd,Pt) is of technological interest because of its relationship to "spin pumping" phenomonology.  Spin pumping, or dynamic exchange coupling, is understood to exert torque on a F layer which is proportional to the time derivative of magnetization, $\partial \mathbf{M}/\partial t$, just as static coupling exerts torque proportional to $\mathbf{M}$.  The dynamic exchange coupling can exert an effective field on $i$ which is $\pi/2$ out of temporal phase with the magnetization of $j$\cite{woltersdorfPRL07} in F$_i$/N/F$_j$, and increases the interface-related Gilbert damping significantly in the F/Cu/N$_2$=(Pd,Pt) system\cite{miz3,ghoshAPL2011}.  \u{S}im\'{a}nek and Heinrich\cite{simanekPRB03} have suggested that static and dynamic exchange coupling share a common origin in the real and imaginary parts, respectively, of the spin correlation function (susceptibility).   However, no evidence for static exchange coupling between $F$ and $N_2$ through $N_1$ has been presented before.

In this paper, we report direct evidence of static, indirect exchange coupling acting on Pd moments in a F/N/Pd superlattice.  Using Pd $L$-edge x-ray magnetic circular dichroism (XMCD), we find sizeable average magnetic moments in Pd, induced by indirect exchange with Ni$_{81}$Fe$_{19}$ (Py) through Cu spacers, roughly 3\% in magnitude of those induced by direct exchange.  The magnitude of the induced moment reflects an effective interlayer exchange coupling energy $J_{ex}$ between Py and Pd which is roughly two orders of magnitude stronger than that present in the Co/Cu/Co system, assuming the bulk paramagnetic susceptibility for Pd.

\section{Experiment}

Two multilayers were prepared by DC magnetron sputtering under computer control.  Base pressure was 2$\times$10$^{-7}\textrm{ Torr}$, working pressure was 1.5 mTorr Ar, and deposition rates were 0.24 nm/s for Cu and $\sim$ 0.1 nm/s for other layers.  Each multilayer was deposited on ion-cleaned Si/SiO$_2$ substrates, with buffer layers of Ta(5nm)/Cu(3nm) to promote (111) fiber texture\cite{nakatanitex}, and was capped with Py(5nm)/Cu(3nm)/Ta(5nm).  Each multilayer consisted of 20 repeats, as substrate/seed/[repeat]$_{20}$/cap.  For the sample in which we investigated Pd moments induced by direct exchange with the Py layers, the repeat unit was [Py(5nm)/Pd(2.5nm)].  We will refer to this sample as "Py/Pd."  For the sample in which we investigated Pd moments induced by indirect exchange with the Py layers, through Cu(3nm), the repeat unit was [Py(5nm)/Cu(3nm)/Pd(2.5nm)/Cu(3nm)].  We will refer to this sample as "Py/Cu/Pd."  The total Pd thickness in each multilayer was thus 50 nm; each Pd layer consists of roughly 11 monolayers, assuming the bulk FCC lattice constant of 3.89 {\AA}.

Pd L-edge XMCD was measured at Beamline ID-12 at the European Synchrotron Radiation Facility (ESRF)\cite{rogalev-chapter}.  The first harmonic of the helical undulator HELIOS II was used to provide circularly polarized X-rays in the energy range between 3.15 and 3.37 keV.  At these energies, the Bragg angle of the Si $<$111$>$ double crystal monochromator is close to the Brewster angle of 45$^{\circ}$, with a consequent reduction of the circular polarization rates from 97\% to about 12.6\% at the Pd $L_3$ edge (3165 eV) and 21.9\% at the Pd $L_2$ edge (3323 eV).  The samples were mounted in a vacuum chamber inserted between poles of an electromagnet generating a magnetic field of 0.6 T.  The incident X-ray beam was parallel/antiparallel to the direction of applied magnetic field, while the angle of incidence at the sample was $\sim$15$^{\circ}$. All spectra were recorded in total fluorescence yield detection mode (TFY).  The sample was kept at room temperature.   XMCD signals were recorded by flipping the direction of magnetic field at each energy point of the spectra. To eliminate any possible experimental artifacts, the XMCD spectra were measured for two opposite helicities of X-rays.  Data acquisition time for each edge was $\sim$ 12 h, thus $\sim$ 48 h for the two samples characterized here.

For quantitative analysis, the XMCD spectra were corrected for incomplete circular polarization and normalized, setting the x-ray absorption above the $L_3$ edge equal to unity and to 0.5 above the $L_2$ absorption edge.
To derive the spin and orbital moments carried by the Pd 4d electrons, the so-called magneto-optical sum rules\cite{tholePRL1992,tholePRL1993} were applied to the normalized XMCD spectra, using

\begin{equation}
<S_z>=\frac{3}{2}(A_3-2A_2)(n_{4d}/\sigma_{tot})-\frac{7}{2}<T_z> \label{sz}
\end{equation}
and
\begin{equation}
<L_z>=2(A_3+A_2)(n_{4d}/\sigma_{tot}) \label{lz}
\end{equation}

where $A_2$ and $A_3$ denotes the integrated XMCD intensities at the $L_2$ and $L_3$ edges,  respectively, $n_{4d}$ is the number of holes in the Pd $4d$ bands, $\sigma_{tot}$ is the total absorption cross-section corresponding to $2p \to 4d$ transitions, and $<T_z>$ is the expectation value of the spin magnetic dipole operator. In the analysis, the contribution of $<T_z>$ was neglected.  Following the well established procedure in ref.\cite{vogelPRB97}, the normalized X-ray absorption cross-section per $4d$-hole, $n_{4d}/\sigma_{tot}$ , was determined by subtracting the Ag-foil L$_{2,3}$ spectra from the experimental Pd L$_{2,3}$ spectra measured on a pure Pd thin film (not shown) and taking the theoretical value for the difference in the 4d holes, 0.92. The same procedure was applied to estimate the number of holes $n_{h,4d}$  on Pd atoms in the samples.

\section{Results}

\begin{figure}
\includegraphics[width=\columnwidth]{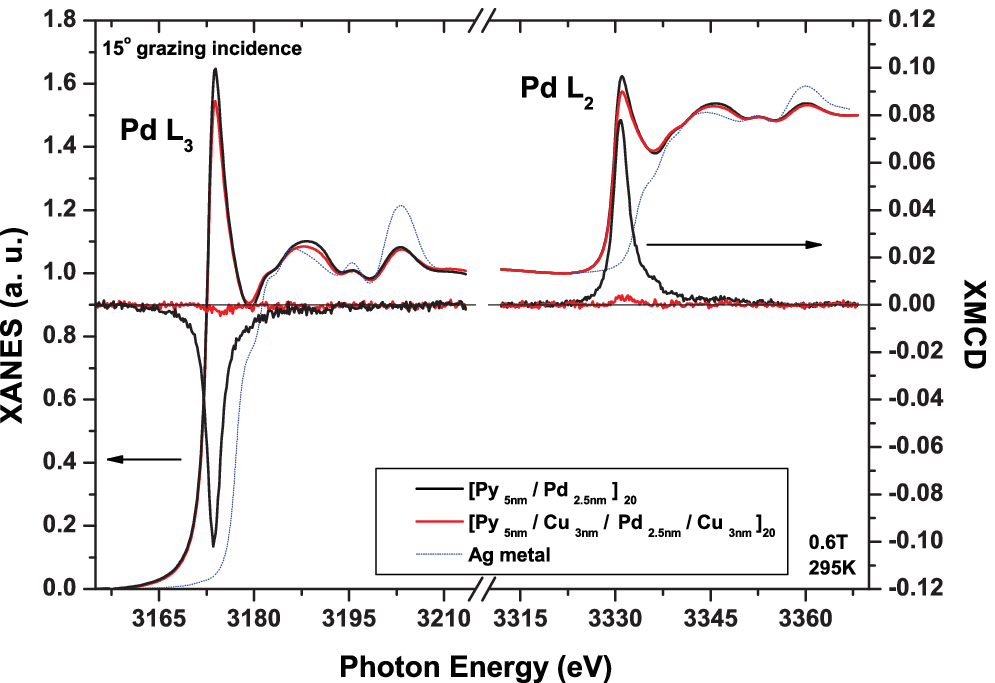}
\centering \caption{X-ray absorption near-edge structure (XANES) and x-ray magnetic circular dichroism (XMCD) for direct exchange-coupled [Py(5nm)/Pd(2.5nm)]$_{20}$ and indirect exchange-coupled [Py(5nm)/Cu(3nm)/Pd(2.5nm)/Cu(3nm)]$_{20}$ samples, with Ag XANES reference.  See text for details.}\label{fig1}\end{figure}

Pd $L_{2,3}$-edge x-ray absorption near-edge structure (XANES) and x-ray magnetic circular dichroism (XMCD) spectra are shown for the Py/Pd and Py/Cu/Pd samples in Figure \ref{fig1}.  A silver-metal XANES spectrum is shown for comparison.  The Pd XANES white lines at 3173 eV (Pd L$_{3}$) and 3331 eV (Pd L$_{2}$) show a $\sim$ 4\% higher absorption for the Py/Pd sample compared with the Py/Cu/Pd sample.  This is attributed to a difference in the number of Pd $4d-$holes, as indicated in Table \ref{table1}, larger in the case of Py/Pd.  The difference can be attributed to charge transfer from the Cu layer into Pd, where $sp$ electrons from Cu fill some of the $4d$ holes near the interface with Pd.  According to the analysis described in the previous section, we estimate a difference in the number of Pd $4d-$holes as $n_{h,4d}=\textrm{1.36}$ for the Py/Pd sample and $n_{h,4d}=\textrm{1.31}$ for the Py/Cu/Pd sample.

Typical negative magnetic circular dichroism (XMCD) at the Pd $L_3$ edge and positive XMCD at the $L_2$ edge are clearly seen in the Py/Pd multilayer, with a maximum XMCD of $\sim -\textrm{10}\%$ at the L$_{3}$ edge and $\sim +\textrm{8}\%$ at L$_{2}$.  According to Eqs \ref{sz} and \ref{lz}, the imbalance in integrated intensities at the $L_{2,3}$ edges indicates a nonzero orbital moment $m_L/m_S$ in Pd.  We estimate $m_L/m_S=\textrm{0.0485}\pm\textrm{0.002}$ for the Pd moments in the direct-exchange coupled sample.

\begin{figure}
\includegraphics[width=\columnwidth]{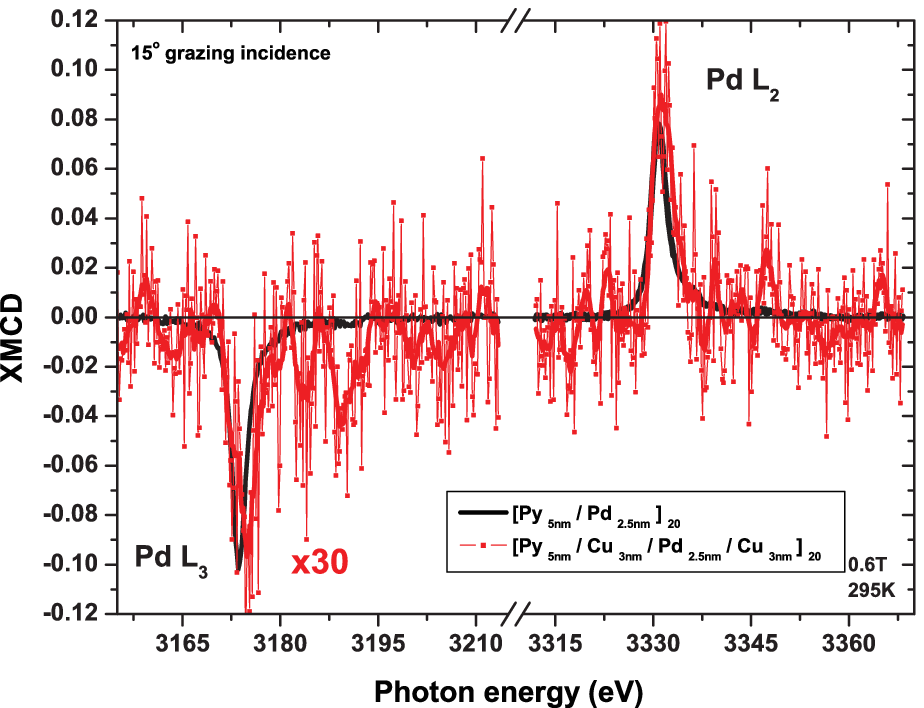}
\centering \caption{X-ray magnetic circular dichroism (XMCD) spectra for [Py(5nm)/Pd(2.5nm)]$_{20}$ and indirect exchange-coupled [Py(5nm)/Cu(3nm)/Pd(2.5nm)/Cu(3nm)]$_{20}$ samples; XMCD for the indirect-exchange coupled sample is magnified by a factor of 30.  Note the transfer of spectral weight from $L_3$ to $L_2$ edges between direct and indirect-exchange coupled Pd.}\label{fig2}\end{figure}

Figure \ref{fig2} shows the central result of the manuscript.  XMCD spectra are compared for the direct-exchange coupled Py/Pd superlattice and the indirect-exchange coupled Py/Cu/Pd superlattice; the data of Figure \ref{fig1} are magnified by a factor of 30 here.  An Pd XMCD signal is clearly present in the Py/Cu/Pd superlattice, roughly 3\% in magnitude that observed for Pd in the Py/Pd superlattice.  It is also apparent in the spectra that the relative weights of the Pd $L_3$ XMCD  and the Pd $L_2$ XMCD are reduced and increased, respectively, through addition of the 3nm Cu spacers.  This reduces the asymmetries between the integrated intensities of $L_{2,3}$ edges, indicating a more nearly pure-spin type moment in Pd for the indirect exchange-coupled sample.

We comment on some more subtle features of the spectra.  There is a slight shift in the Pd $L_{3}$ and $L_{2}$ peaks towards higher energies in the Py/Cu/Pd sample compared with the Py/Pd sample, by $\sim$ 2 eV.  No background has been subtracted from the XANES spectra in the data presented; the shift is present in the raw XMCD data as well.  We regard this to be a real effect, due to changes in the $4d-$band occupation in the multilayer.  Second, there are spectral features at 3187 and 3345 eV in the XMCD, corresponding to the first EXAFS wiggles in the corresponding absorption spectra.  These features are tentatively assigned to magnetic EXAFS, and are more pronounced in the Py/Cu/Pd sample because the XMCD signal is much weaker.

The sum-rules-extracted moments are tabulated in Table \ref{table1}.  We find that the moment induced through direct exchange in the Py/Pd system is quite similar to that found in Fe/Pd multilayers.  Vogel et al\cite{vogelPRB97} found that over 4 ML Pd(100) from the Fe interface, or 0.78 nm of Pd, there exists $\left<m_{Pd}\right>\sim\textrm{0.27}\mu_B/\textrm{at}$. Thus for the 0.78 nm of Pd next to Py on either side of a Py/Pd(2.5nm)/Py structure, 1.6 nm total, we expect a moment of that magnitude, diluted by Pd layers with near-zero moment in the interior of the Pd layer.  This gives an estimate of $\sim\textrm{0.17}\mu_{B}$, larger by $\sim$ 40\% than that observed here for Ni$_{81}$Fe$_{19}$/Pd(111).  Taking an induced spin moment of 0.012 $\mu_{B}/\textrm{at}$/ 7 T as in \cite{rogalev-chapter}, the effective field from direct exchange is roughly 70 T over the 2.5 nm Pd layer.  The $m_{L}/m_{S}$ value agrees well with 0.045 observed for Fe(10ML)/Pd(14ML)\cite{vogelPRB97}.

The moment induced in Pd by indirect exchange through Cu is weaker by a factor of 30, but clearly visible in the spectrum.  It corresponds to an effective field acting on Pd of $\sim$ 2 T.  Sum-rules analysis indicates that $m_L/m_S=$2.8\%, with a reduced fraction of the moment coming from the orbital component.  The greater spin-type component is consistent with RKKY-type exchange through Cu, which possesses induced spin moments with negligible orbital components\cite{stohr-cocuco}, acting on Pd.  Error bars are large on the estimate due to the small magnitude of the moment, but the relative increase of the $L_2$ dichroism for the indirect exchange sample is clear in Figure \ref{fig2}.

\begin{table}
\begin{tabular}{|c|c|c|c|}
  \hline
  Sample & m$^{tot}_{Pd}$ ($\mu_B/\textrm{at}$) & $m_L/m_S$ & $n_{h,4d}/\textrm{at}$ \\
  \hline
  Py/Pd & $+\textrm{0.1160}\pm\textrm{0.0007}$ & $\textrm{0.0485}\pm\textrm{0.002}$ & 1.36 \\
  Py/Cu/Pd & $+\textrm{0.0036}\pm\textrm{0.0007}$ & $\textrm{0.028}\pm\textrm{0.08}$ & 1.31 \\
  \hline
\end{tabular}
\caption{Pd-site total magnetic moments $m^{tot}$, ratio of orbital to spin moments $m_{L}/m_S$, and number of $d-$holes extracted from Pd $L-edge$ XMCD measurements.}
\label{table1}
\end{table}

\section{Discussion}

In the past, interlayer exchange coupling (IEC) energies have been estimated through effective fields acting on (fixed magnitude) ferromagnetic moments $M_s$.  In $cgs$ units, the energy splitting is $\Delta E = - J_{ex} / t_{FM}$, with effective field $\mathbf{H_{ex}}= -\partial E / \partial\mathbf{M} = J_{ex}/(M_s\:t_{FM})$.  For Co/Cu(100), $t_{Cu}\simeq \textrm{2.1 nm}$, Qiu et al\cite{qiuPRB92} determined an interfacial exchange coupling parameter of $J_{ex}=\textrm{0.05 erg/cm}^2$ from magnetooptical Kerr effect (MOKE) measurements of wedged trilayers, or a saturation field of $\sim$ 80 G, after the second antiferromagnetic maximum.

In our experiment, the interfacial interlayer exchange coupling creates a magnetic moment in Pd, rather than simply shifting its saturation field.  We will infer an exchange energy from the induced moment assuming the bulk paramagnetic susceptibility of Pd.  The induced moment $M_{Pd}$ from an exchange energy splitting $\Delta E_{ex}$ is $M_{Pd}=2 F\mu_B N_0\Delta E_{ex}$, where $F$ is the Stoner factor, $\mu_B$ is the Bohr magneton, and $N_0$ is the density of states at the Fermi energy of Pd.  We estimate the total IEC exchange energy acting on the Pd, to generate the moment, as

\begin{equation}
\Delta E_{ex} = {M_{Pd}\over 2 F\mu_B N_0}
\end{equation}

We find that the moment induced through indirect exchange is significantly larger than that expected from interlayer exchange coupling and the bulk paramagnetic susceptibility.  Taking the observed $M_{Pd} = \textrm{0.0038}\: \mu_{B}/\textrm{at}$, $N_0\simeq \textrm{1.3}\textrm{/(eV}\cdot\textrm{at}\cdot{\textrm{spin)}}$\cite{l-b-eband}, and the bulk $F\simeq\textrm{9.2}$ gives $\Delta E_{ex}=\textrm{150}\:\mu\textrm{eV/at}$.  An atomic volume of $V=a^3/4=\textrm{14.7 \AA}^3/\textrm{at}$ can be applied to estimate the interfacial exchange coupling by $J_{ex}=+\Delta E\:t_{Pd}/V$, given an assumed value of $F$.  Note that the estimate of $J_{ex}$ would not change if we assumed that the Pd magnetic moment is generated only in a few interface layers: the experimental moment $M$ and the conversion $J_{ex}/\Delta E$ would change by the same proportionality.

In the experiment, we can probe only the $F\cdot J_{ex}$ product.  Assuming the bulk paramagnetic susceptibility enhancement factor $F$, the interfacial indirect exchange coupling energy, $J_{ex}=\textrm{+4.3 erg/cm}^2$, will be two orders of magnitude (215 x) stronger than that seen in \cite{qiuPRB92} for epitaxial Co/Cu/Co(001) at 3nm Cu thickness ($\sim$0.02 erg/cm$^{2}$).  Alternatively, assuming an identical value of exchange coupling yields an estimate of susceptibility enhancement factor $F\sim\textrm{2000}$, two orders of magnitude above the bulk susceptibility.

\begin{figure}
\includegraphics[width=\columnwidth]{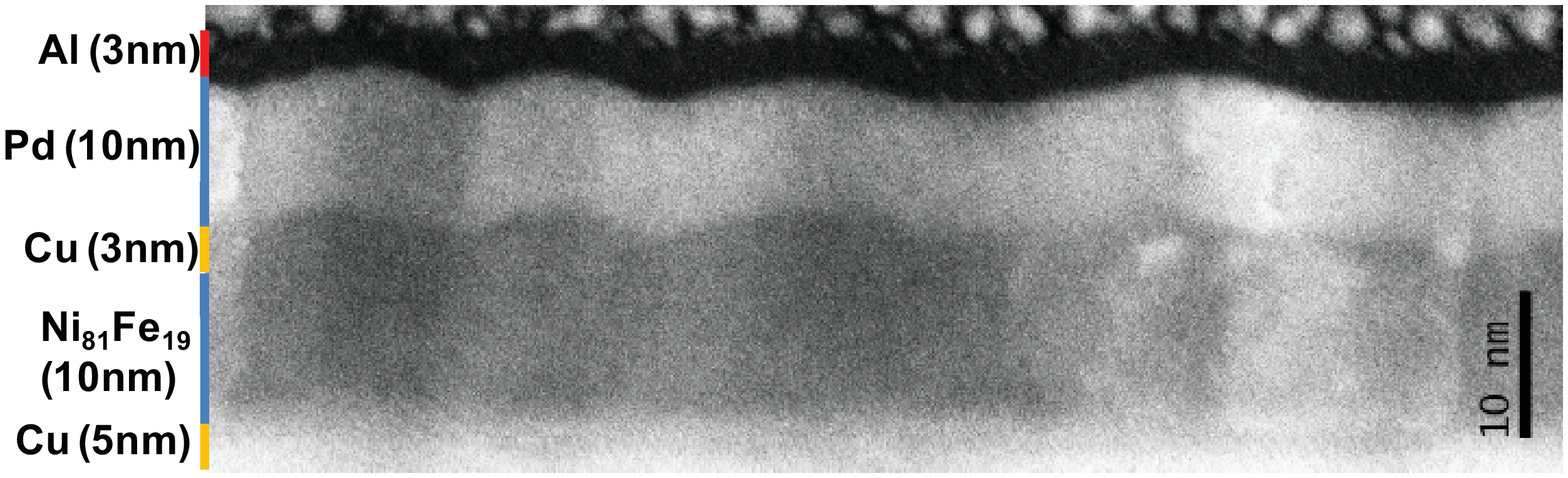}
\centering \caption{Bright-field cross-sectional TEM image for seed/Ni$_{81}$Fe$_{19}$(10nm)/Cu(3nm)/Pd(10nm)/Al(3nm).}\label{fig3}\end{figure}

Previously, Ir magnetic moments of $\sim$ 0.01 $\mu_{B}/\textrm{at}$ have been observed in a [Co(0.4)/ Pt(1.1)/ Ir(0.8)/ Pt(1.1nm)]$_{25}$ heterostructure\cite{schutzJAP93}.  This is an "existence proof" that magnetic order can be induced in one bulk paramagnet through another bulk paramagnet.  However, adjusted for thickness, the induced Ir moment is significantly weaker than our observation for Pd in [Py(5nm)/ Cu(3nm)/ Pd(2.5nm)/ Cu(3nm)]$_{20}$: the moment-Pd (-Ir) thickness product is comparable, and a variation as $1/t_{Pt}^2$ would reduce the Ir moment by an order of magnitude at 3 nm.  Moreover, the mechanism may be different, as it is also less clear that the induced Ir moment relates to RKKY through the Pt, given the absence of oscillatory coupling seen in Co/Pt.\cite{liPRB94}

A possible interpretation for the large exchange coupling $J_{ex}$ in the Py/Cu/Pd structure is that the measured value is dominated by direct contact of Py and Pd through discontinuities ("pinholes") in the 3nm Cu layers.  However, we do not believe that this is likely.  First, a cross-sectional TEM image (Figure \ref{fig3}) for a single, thicker Pd(10nm) layer, seeded similarly, shows continuous coverage, with Cu/Pd interface roughness conformal with the top surface roughness.  This reproduces findings of older TEM investigations which show conformal coverage of Pd or Pt in multiple-repeat (x20) heterostructures with Pd as thin as 1 nm\cite{berteroJMMM94,shinjoJMMM93}.  Second, ferromagnetic coupling between identically seeded $F_1$/Cu(3nm)/$F_2$ structures is $<$ 20 Oe for 8 nm Co layers in ferromagnetic resonance (FMR)\cite{ghoshPRL12}; this places an upper bound of $J_{coup}\sim\textrm{0.02 erg/cm}^2$, likely dominated by dipolar (Neel) coupling.  Third, direct exchange would require that roughly one Pd atom out of 30 is in direct contact with the Py layer; given the thickness of the Cu layer, $\sim$ 15 monolayers, this is an unrealistically large value.  Finally, the XMCD shape (L/S ratio) is quite different for the Py/Pd and Py/Cu/Pd structures.  This signal is sensitive to only the magnetically active Pd atoms in the structure.  The lower orbital moment component observed for the Py/Cu/Pd structure is consistent with (spin-mediated) indirect exchange through Cu.

Finally, we note that dramatically enhanced susceptibilities $F$ have been reported for ultrathin Pd before.  Brodsky and Freeman\cite{brodskyPRL80} observed susceptibility enhancements up to a factor of 500 over bulk values in Au/Pd/Au, attributed to lattice distortion of the ultrathin Pd.  There was some discussion in this work regarding a possible role of Fe impurities or entrapped O$_2$.  In our experiments, the XMCD technique provides confidence that the enhanced moments are located on Pd sites.

\section{Summary}

We have shown that Pd, with its greatest tendency towards ferromagnetism of all paramagnets, can have significant magnetic order induced through long-range indirect exchange through 3 nm of Cu.  Estimated values for the exchange coupling strength between ferromagnetic permalloy (Py) and Pd are two orders of magnitude larger than those found in the Co/Cu/Co system.

Our results have some implications for the understanding of damping enhancement through spin pumping, or dynamic exchange coupling\cite{heinrich-2003}.  Static and dynamic interlayer exchange couplings are not typically calculated on equal footing.  Widely used models for the damping enhancement\cite{brataasPRL00,bauerPRB05} assume that electron momenta are randomized at the interfaces.  This implicitly assumes near-zero static exchange coupling, justified to date because the Cu thicknesses typically investigated in spin pumping studies, $>$ 3 nm, are beyond the range where RKKY is strong, and since no RKKY mechanism has been obvious in heterostructures of F/Cu/$4d, 5d$ "spin sink" metals.

Our experimental results call both these assumptions into question for F/Cu/Pd: the RKKY mechanism is evidently nonnegligible at 3 nm and stronger than expected from studies with two F layers.  It is possible that where the IEC is strong, it has a measurable effect on the damping enhancement $\Delta\alpha$: Costa et al\cite{costaPRB06} predict that a component of $\Delta\alpha$--nearly half the asymptotic value for Co/Cu/Co(100), for $t_{Cu}$ of a few monolayers--does indeed oscillate with the same period as the IEC.  Because of the complicated nature of the FMR spectra for strongly coupled F layers, it has been difficult to investigate the predicted behavior.  Our results suggest that the Py/Cu/Pd system, with a single FMR mode but evidently strong coupling, is suitable for studies of the interplay between static and dynamic coupling.

\section{Acknowledgements}

We acknowledge support from the U.S. NSF-ECCS-0925829.
\printfigures


\begin{thebibliography}{32}%
\makeatletter
\providecommand \@ifxundefined [1]{%
 \@ifx{#1\undefined}
}%
\providecommand \@ifnum [1]{%
 \ifnum #1\expandafter \@firstoftwo
 \else \expandafter \@secondoftwo
 \fi
}%
\providecommand \@ifx [1]{%
 \ifx #1\expandafter \@firstoftwo
 \else \expandafter \@secondoftwo
 \fi
}%
\providecommand \natexlab [1]{#1}%
\providecommand \enquote  [1]{``#1''}%
\providecommand \bibnamefont  [1]{#1}%
\providecommand \bibfnamefont [1]{#1}%
\providecommand \citenamefont [1]{#1}%
\providecommand \href@noop [0]{\@secondoftwo}%
\providecommand \href [0]{\begingroup \@sanitize@url \@href}%
\providecommand \@href[1]{\@@startlink{#1}\@@href}%
\providecommand \@@href[1]{\endgroup#1\@@endlink}%
\providecommand \@sanitize@url [0]{\catcode `\\12\catcode `\$12\catcode
  `\&12\catcode `\#12\catcode `\^12\catcode `\_12\catcode `\%12\relax}%
\providecommand \@@startlink[1]{}%
\providecommand \@@endlink[0]{}%
\providecommand \url  [0]{\begingroup\@sanitize@url \@url }%
\providecommand \@url [1]{\endgroup\@href {#1}{\urlprefix }}%
\providecommand \urlprefix  [0]{URL }%
\providecommand \Eprint [0]{\href }%
\providecommand \doibase [0]{http://dx.doi.org/}%
\providecommand \selectlanguage [0]{\@gobble}%
\providecommand \bibinfo  [0]{\@secondoftwo}%
\providecommand \bibfield  [0]{\@secondoftwo}%
\providecommand \translation [1]{[#1]}%
\providecommand \BibitemOpen [0]{}%
\providecommand \bibitemStop [0]{}%
\providecommand \bibitemNoStop [0]{.\EOS\space}%
\providecommand \EOS [0]{\spacefactor3000\relax}%
\providecommand \BibitemShut  [1]{\csname bibitem#1\endcsname}%
\let\auto@bib@innerbib\@empty
%</preamble>
\bibitem [{l-b(1990{\natexlab{a}})}]{l-b-4d5d}%
  \BibitemOpen
  \enquote {\bibinfo {title} {Landolt-bornstein tables},}\ Chap.\ \bibinfo
  {chapter} {1.3.1: Introduction to the paramagnetism of 4d and 5d transition metals}, p.\ \bibinfo
  {pages} {492},\ in\  \cite{l-b-eband-book} (\bibinfo {year}
  {1990}{\natexlab{a}})\BibitemShut {NoStop}%
\bibitem [{\citenamefont {Stoner}(1936)}]{stonerPRS36}%
  \BibitemOpen
  \bibfield  {author} {\bibinfo {author} {\bibfnamefont {E.}~\bibnamefont
  {Stoner}},\ }\href@noop {} {\bibfield  {journal} {\bibinfo  {journal}
  {Proceedings of the Royal Society of London A}\ }\textbf {\bibinfo {volume}
  {154}},\ \bibinfo {pages} {656} (\bibinfo {year} {1936})}\BibitemShut
  {NoStop}%
\bibitem [{\citenamefont {Shimizu}(1981)}]{shimizuRPP81}%
  \BibitemOpen
  \bibfield  {author} {\bibinfo {author} {\bibfnamefont {M.}~\bibnamefont
  {Shimizu}},\ }\href@noop {} {\bibfield  {journal} {\bibinfo  {journal}
  {Reports on Progress in Physics}\ }\textbf {\bibinfo {volume} {44}},\
  \bibinfo {pages} {329} (\bibinfo {year} {1981})}\BibitemShut {NoStop}%
\bibitem [{\citenamefont {Schindler}\ and\ \citenamefont
  {Rice}(1967)}]{schindlerPR67}%
  \BibitemOpen
  \bibfield  {author} {\bibinfo {author} {\bibfnamefont {A.}~\bibnamefont
  {Schindler}}\ and\ \bibinfo {author} {\bibfnamefont {M.}~\bibnamefont
  {Rice}},\ }\href@noop {} {\bibfield  {journal} {\bibinfo  {journal} {Physical
  Review}\ }\textbf {\bibinfo {volume} {164}},\ \bibinfo {pages} {759}
  (\bibinfo {year} {1967})}\BibitemShut {NoStop}%
\bibitem [{\citenamefont {Berk}\ and\ \citenamefont
  {Schrieffer}(1966)}]{berkPRL66}%
  \BibitemOpen
  \bibfield  {author} {\bibinfo {author} {\bibfnamefont {N.}~\bibnamefont
  {Berk}}\ and\ \bibinfo {author} {\bibfnamefont {J.}~\bibnamefont
  {Schrieffer}},\ }\href@noop {} {\bibfield  {journal} {\bibinfo  {journal}
  {Phys. Rev. Lett.}\ }\textbf {\bibinfo {volume} {17}},\ \bibinfo {pages}
  {433} (\bibinfo {year} {1966})}\BibitemShut {NoStop}%
\bibitem [{\citenamefont {Hong}\ \emph {et~al.}(1988)\citenamefont {Hong},
  \citenamefont {Furdyna},\ and\ \citenamefont {Freeman}}]{hongJAP88}%
  \BibitemOpen
  \bibfield  {author} {\bibinfo {author} {\bibfnamefont {J.}~\bibnamefont
  {Hong}}, \bibinfo {author} {\bibfnamefont {J.}~\bibnamefont {Furdyna}}, \
  and\ \bibinfo {author} {\bibfnamefont {A.}~\bibnamefont {Freeman}},\
  }\href@noop {} {\bibfield  {journal} {\bibinfo  {journal} {Journal of Applied
  Physics}\ }\textbf {\bibinfo {volume} {63}},\ \bibinfo {pages} {3655}
  (\bibinfo {year} {1988})}\BibitemShut {NoStop}%
\bibitem [{\citenamefont {Brodsky}\ and\ \citenamefont
  {Freeman}(1980)}]{brodskyPRL80}%
  \BibitemOpen
  \bibfield  {author} {\bibinfo {author} {\bibfnamefont {M.~B.}\ \bibnamefont
  {Brodsky}}\ and\ \bibinfo {author} {\bibfnamefont {A.~J.}\ \bibnamefont
  {Freeman}},\ }\href {\doibase 10.1103/PhysRevLett.45.133} {\bibfield
  {journal} {\bibinfo  {journal} {Phys. Rev. Lett.}\ }\textbf {\bibinfo
  {volume} {45}},\ \bibinfo {pages} {133} (\bibinfo {year} {1980})}\BibitemShut
  {NoStop}%
\bibitem [{\citenamefont {Cable}\ \emph {et~al.}(1962)\citenamefont {Cable},
  \citenamefont {Wollan}, \citenamefont {Koehler},\ and\ \citenamefont
  {Wilkinson}}]{cableJAP62}%
  \BibitemOpen
  \bibfield  {author} {\bibinfo {author} {\bibfnamefont {J.~W.}\ \bibnamefont
  {Cable}}, \bibinfo {author} {\bibfnamefont {E.~O.}\ \bibnamefont {Wollan}},
  \bibinfo {author} {\bibfnamefont {W.~C.}\ \bibnamefont {Koehler}}, \ and\
  \bibinfo {author} {\bibfnamefont {M.~K.}\ \bibnamefont {Wilkinson}},\
  }\href@noop {} {\bibfield  {journal} {\bibinfo  {journal} {Journal of Applied
  Physics}\ }\textbf {\bibinfo {volume} {33}},\ \bibinfo {pages} {1340}
  (\bibinfo {year} {1962})}\BibitemShut {NoStop}%
\bibitem [{\citenamefont {Thole}\ \emph {et~al.}(1992)\citenamefont {Thole},
  \citenamefont {Carra}, \citenamefont {Sette},\ and\ \citenamefont {van~der
  Laan}}]{tholePRL1992}%
  \BibitemOpen
  \bibfield  {author} {\bibinfo {author} {\bibfnamefont {B.}~\bibnamefont
  {Thole}}, \bibinfo {author} {\bibfnamefont {P.}~\bibnamefont {Carra}},
  \bibinfo {author} {\bibfnamefont {F.}~\bibnamefont {Sette}}, \ and\ \bibinfo
  {author} {\bibfnamefont {G.}~\bibnamefont {van~der Laan}},\ }\href@noop {}
  {\bibfield  {journal} {\bibinfo  {journal} {Physical Review Letters}\
  }\textbf {\bibinfo {volume} {68}},\ \bibinfo {pages} {1943} (\bibinfo {year}
  {1992})}\BibitemShut {NoStop}%
\bibitem [{\citenamefont {Vogel}\ \emph {et~al.}(1997)\citenamefont {Vogel},
  \citenamefont {Fontaine}, \citenamefont {Cros}, \citenamefont {Petroff},
  \citenamefont {Kappler}, \citenamefont {Krill}, \citenamefont {Rogalev},\
  and\ \citenamefont {Goulon}}]{vogelPRB97}%
  \BibitemOpen
  \bibfield  {author} {\bibinfo {author} {\bibfnamefont {J.}~\bibnamefont
  {Vogel}}, \bibinfo {author} {\bibfnamefont {A.}~\bibnamefont {Fontaine}},
  \bibinfo {author} {\bibfnamefont {V.}~\bibnamefont {Cros}}, \bibinfo {author}
  {\bibfnamefont {F.}~\bibnamefont {Petroff}}, \bibinfo {author} {\bibfnamefont
  {J.-P.}\ \bibnamefont {Kappler}}, \bibinfo {author} {\bibfnamefont
  {G.}~\bibnamefont {Krill}}, \bibinfo {author} {\bibfnamefont
  {A.}~\bibnamefont {Rogalev}}, \ and\ \bibinfo {author} {\bibfnamefont
  {J.}~\bibnamefont {Goulon}},\ }\href@noop {} {\bibfield  {journal} {\bibinfo
  {journal} {Physical Review B (Condensed Matter)}\ }\textbf {\bibinfo {volume}
  {55}},\ \bibinfo {pages} {3663 } (\bibinfo {year} {1997})}\BibitemShut
  {NoStop}%
\bibitem [{\citenamefont {Ruderman}\ and\ \citenamefont
  {Kittel}(1954)}]{rudermanPR54}%
  \BibitemOpen
  \bibfield  {author} {\bibinfo {author} {\bibfnamefont {M.}~\bibnamefont
  {Ruderman}}\ and\ \bibinfo {author} {\bibfnamefont {C.}~\bibnamefont
  {Kittel}},\ }\href@noop {} {\bibfield  {journal} {\bibinfo  {journal}
  {Physical Review}\ }\textbf {\bibinfo {volume} {96}},\ \bibinfo {pages} {99}
  (\bibinfo {year} {1954})}\BibitemShut {NoStop}%
\bibitem [{\citenamefont {Parkin}(1991)}]{parkincoupling}%
  \BibitemOpen
  \bibfield  {author} {\bibinfo {author} {\bibfnamefont {S.}~\bibnamefont
  {Parkin}},\ }\href@noop {} {\bibfield  {journal} {\bibinfo  {journal}
  {Physical Review Letters}\ }\textbf {\bibinfo {volume} {67}},\ \bibinfo
  {pages} {3598} (\bibinfo {year} {1991})}\BibitemShut {NoStop}%
\bibitem [{\citenamefont {Li}\ \emph {et~al.}(2004)\citenamefont {Li},
  \citenamefont {Ma}, \citenamefont {Peng}, \citenamefont {Zhao}, \citenamefont
  {Mei}, \citenamefont {Gu}, \citenamefont {Chai}, \citenamefont {Mai},
  \citenamefont {Shen}, \citenamefont {Liu},\ and\ \citenamefont
  {Dai}}]{liPRB94}%
  \BibitemOpen
  \bibfield  {author} {\bibinfo {author} {\bibfnamefont {M.}~\bibnamefont
  {Li}}, \bibinfo {author} {\bibfnamefont {X.}~\bibnamefont {Ma}}, \bibinfo
  {author} {\bibfnamefont {C.}~\bibnamefont {Peng}}, \bibinfo {author}
  {\bibfnamefont {G.}~\bibnamefont {Zhao}}, \bibinfo {author} {\bibfnamefont
  {I.}~\bibnamefont {Mei}}, \bibinfo {author} {\bibfnamefont {Y.}~\bibnamefont
  {Gu}}, \bibinfo {author} {\bibfnamefont {W.}~\bibnamefont {Chai}}, \bibinfo
  {author} {\bibfnamefont {Z.}~\bibnamefont {Mai}}, \bibinfo {author}
  {\bibfnamefont {B.}~\bibnamefont {Shen}}, \bibinfo {author} {\bibfnamefont
  {Y.}~\bibnamefont {Liu}}, \ and\ \bibinfo {author} {\bibfnamefont
  {D.}~\bibnamefont {Dai}},\ }\href@noop {} {\bibfield  {journal} {\bibinfo
  {journal} {Physical Review B}\ }\textbf {\bibinfo {volume} {50}},\ \bibinfo
  {pages} {10323} (\bibinfo {year} {1994})}\BibitemShut {NoStop}%
\bibitem [{\citenamefont {Woltersdorf}\ \emph {et~al.}(2007)\citenamefont
  {Woltersdorf}, \citenamefont {Mosendz}, \citenamefont {Heinrich},\ and\
  \citenamefont {Back}}]{woltersdorfPRL07}%
  \BibitemOpen
  \bibfield  {author} {\bibinfo {author} {\bibfnamefont {G.}~\bibnamefont
  {Woltersdorf}}, \bibinfo {author} {\bibfnamefont {O.}~\bibnamefont
  {Mosendz}}, \bibinfo {author} {\bibfnamefont {B.}~\bibnamefont {Heinrich}}, \
  and\ \bibinfo {author} {\bibfnamefont {C.~H.}\ \bibnamefont {Back}},\
  }\href@noop {} {\bibfield  {journal} {\bibinfo  {journal} {Physical Review
  Letters}\ }\textbf {\bibinfo {volume} {99}},\ \bibinfo {pages} {246603}
  (\bibinfo {year} {2007})}\BibitemShut {NoStop}%
\bibitem [{\citenamefont {Mizukami}\ \emph {et~al.}(2002)\citenamefont
  {Mizukami}, \citenamefont {Ando},\ and\ \citenamefont {Miyazaki}}]{miz3}%
  \BibitemOpen
  \bibfield  {author} {\bibinfo {author} {\bibfnamefont {S.}~\bibnamefont
  {Mizukami}}, \bibinfo {author} {\bibfnamefont {Y.}~\bibnamefont {Ando}}, \
  and\ \bibinfo {author} {\bibfnamefont {T.}~\bibnamefont {Miyazaki}},\ }\href
  {\doibase 10.1103/PhysRevB.66.104413} {\bibfield  {journal} {\bibinfo
  {journal} {Phys. Rev. B}\ }\textbf {\bibinfo {volume} {66}},\ \bibinfo
  {pages} {104413} (\bibinfo {year} {2002})}\BibitemShut {NoStop}%
\bibitem [{\citenamefont {Ghosh}\ \emph {et~al.}(2011)\citenamefont {Ghosh},
  \citenamefont {Sierra}, \citenamefont {Auffret}, \citenamefont {Ebels},\ and\
  \citenamefont {Bailey}}]{ghoshAPL2011}%
  \BibitemOpen
  \bibfield  {author} {\bibinfo {author} {\bibfnamefont {A.}~\bibnamefont
  {Ghosh}}, \bibinfo {author} {\bibfnamefont {J.~F.}\ \bibnamefont {Sierra}},
  \bibinfo {author} {\bibfnamefont {S.}~\bibnamefont {Auffret}}, \bibinfo
  {author} {\bibfnamefont {U.}~\bibnamefont {Ebels}}, \ and\ \bibinfo {author}
  {\bibfnamefont {W.~E.}\ \bibnamefont {Bailey}},\ }\href {\doibase
  10.1063/1.3551729} {\bibfield  {journal} {\bibinfo  {journal} {Applied
  Physics Letters}\ }\textbf {\bibinfo {volume} {98}},\ \bibinfo {eid} {052508}
  (\bibinfo {year} {2011})}\BibitemShut {NoStop}%
\bibitem [{\citenamefont {Simanek}\ and\ \citenamefont
  {Heinrich}(2003)}]{simanekPRB03}%
  \BibitemOpen
  \bibfield  {author} {\bibinfo {author} {\bibfnamefont {E.}~\bibnamefont
  {Simanek}}\ and\ \bibinfo {author} {\bibfnamefont {B.}~\bibnamefont
  {Heinrich}},\ }\href@noop {} {\bibfield  {journal} {\bibinfo  {journal}
  {Physical Review B (Condensed Matter)}\ }\textbf {\bibinfo {volume} {67}},\
  \bibinfo {pages} {144418} (\bibinfo {year} {2003})}\BibitemShut {NoStop}%
\bibitem [{\citenamefont {R.~Nakatani}\ and\ \citenamefont
  {Sugita}(1994)}]{nakatanitex}%
  \BibitemOpen
  \bibfield  {author} {\bibinfo {author} {\bibfnamefont {S.~N.}\ \bibnamefont
  {R.~Nakatani}, \bibfnamefont {K.~Hoshino}}\ and\ \bibinfo {author}
  {\bibfnamefont {Y.}~\bibnamefont {Sugita}},\ }\href@noop {} {\bibfield
  {journal} {\bibinfo  {journal} {Japanese Journal of Applied Physics}\
  }\textbf {\bibinfo {volume} {33}},\ \bibinfo {pages} {133} (\bibinfo {year}
  {1994})}\BibitemShut {NoStop}%
\bibitem [{\citenamefont {Rogalev}\ \emph {et~al.}(2006)\citenamefont
  {Rogalev}, \citenamefont {Wilhelm}, \citenamefont {Jaouen}, \citenamefont
  {Goulon},\ and\ \citenamefont {Kappler}}]{rogalev-chapter}%
  \BibitemOpen
  \bibfield  {author} {\bibinfo {author} {\bibfnamefont {A.}~\bibnamefont
  {Rogalev}}, \bibinfo {author} {\bibfnamefont {F.}~\bibnamefont {Wilhelm}},
  \bibinfo {author} {\bibfnamefont {N.}~\bibnamefont {Jaouen}}, \bibinfo
  {author} {\bibfnamefont {J.}~\bibnamefont {Goulon}}, \ and\ \bibinfo {author}
  {\bibfnamefont {J.-P.}\ \bibnamefont {Kappler}},\ }\enquote {\bibinfo {title}
  {X-ray magnetic circular dichroism: historical perspective and recent
  highlights},}\ \ (\bibinfo {address} {Berlin, Germany},\ \bibinfo {year}
  {2006})\ pp.\ \bibinfo {pages} {71 -- 93}\BibitemShut {NoStop}%
\bibitem [{\citenamefont {Carra}\ \emph {et~al.}(1993)\citenamefont {Carra},
  \citenamefont {Thole}, \citenamefont {Altarelli},\ and\ \citenamefont
  {Wang}}]{tholePRL1993}%
  \BibitemOpen
  \bibfield  {author} {\bibinfo {author} {\bibfnamefont {P.}~\bibnamefont
  {Carra}}, \bibinfo {author} {\bibfnamefont {B.}~\bibnamefont {Thole}},
  \bibinfo {author} {\bibfnamefont {M.}~\bibnamefont {Altarelli}}, \ and\
  \bibinfo {author} {\bibfnamefont {X.}~\bibnamefont {Wang}},\ }\href@noop {}
  {\bibfield  {journal} {\bibinfo  {journal} {Physical Review Letters}\
  }\textbf {\bibinfo {volume} {70}},\ \bibinfo {pages} {694} (\bibinfo {year}
  {1993})}\BibitemShut {NoStop}%
\bibitem [{\citenamefont {Samant}\ \emph {et~al.}(1994)\citenamefont {Samant},
  \citenamefont {St{\"o}hr}, \citenamefont {Parkin}, \citenamefont {Held},
  \citenamefont {Hermsmeier}, \citenamefont {Herman}, \citenamefont {van
  Schilfgaarde}, \citenamefont {Duda}, \citenamefont {Mancini}, \citenamefont
  {Wassdahl},\ and\ \citenamefont {Nakajima}}]{stohr-cocuco}%
  \BibitemOpen
  \bibfield  {author} {\bibinfo {author} {\bibfnamefont {M.}~\bibnamefont
  {Samant}}, \bibinfo {author} {\bibfnamefont {J.}~\bibnamefont {St{\"o}hr}},
  \bibinfo {author} {\bibfnamefont {S.}~\bibnamefont {Parkin}}, \bibinfo
  {author} {\bibfnamefont {G.}~\bibnamefont {Held}}, \bibinfo {author}
  {\bibfnamefont {B.}~\bibnamefont {Hermsmeier}}, \bibinfo {author}
  {\bibfnamefont {F.}~\bibnamefont {Herman}}, \bibinfo {author} {\bibfnamefont
  {M.}~\bibnamefont {van Schilfgaarde}}, \bibinfo {author} {\bibfnamefont
  {L.-C.}\ \bibnamefont {Duda}}, \bibinfo {author} {\bibfnamefont
  {D.}~\bibnamefont {Mancini}}, \bibinfo {author} {\bibfnamefont
  {N.}~\bibnamefont {Wassdahl}}, \ and\ \bibinfo {author} {\bibfnamefont
  {R.}~\bibnamefont {Nakajima}},\ }\href@noop {} {\bibfield  {journal}
  {\bibinfo  {journal} {Physical Review Letters}\ }\textbf {\bibinfo {volume}
  {72}},\ \bibinfo {pages} {1112} (\bibinfo {year} {1994})}\BibitemShut
  {NoStop}%
\bibitem [{\citenamefont {Qiu}\ \emph {et~al.}(1992)\citenamefont {Qiu},
  \citenamefont {Pearson},\ and\ \citenamefont {Bader}}]{qiuPRB92}%
  \BibitemOpen
  \bibfield  {author} {\bibinfo {author} {\bibfnamefont {Z.}~\bibnamefont
  {Qiu}}, \bibinfo {author} {\bibfnamefont {J.}~\bibnamefont {Pearson}}, \ and\
  \bibinfo {author} {\bibfnamefont {S.}~\bibnamefont {Bader}},\ }\href@noop {}
  {\bibfield  {journal} {\bibinfo  {journal} {Physical Review B (Condensed
  Matter)}\ }\textbf {\bibinfo {volume} {46}},\ \bibinfo {pages} {8659}
  (\bibinfo {year} {1992})}\BibitemShut {NoStop}%
\bibitem [{l-b(1990{\natexlab{b}})}]{l-b-eband}%
  \BibitemOpen
  \enquote {\bibinfo {title} {Landolt-bornstein tables},}\ Chap.\ \bibinfo
  {chapter} {III-13: Metals: Phonon States, Electron States, and Fermi
  Surfaces}, pp.\ \bibinfo {pages} {100--101},\ in\  \cite{l-b-eband-book}
  (\bibinfo {year} {1990}{\natexlab{b}})\BibitemShut {NoStop}%
\bibitem [{\citenamefont {Sch\"{u}tz}\ \emph {et~al.}(1993)\citenamefont
  {Sch\"{u}tz}, \citenamefont {St\"{a}hler}, \citenamefont {Kn\"{u}lle},
  \citenamefont {Fischer}, \citenamefont {Parkin},\ and\ \citenamefont
  {Ebert}}]{schutzJAP93}%
  \BibitemOpen
  \bibfield  {author} {\bibinfo {author} {\bibfnamefont {G.}~\bibnamefont
  {Sch\"{u}tz}}, \bibinfo {author} {\bibfnamefont {S.}~\bibnamefont
  {St\"{a}hler}}, \bibinfo {author} {\bibfnamefont {M.}~\bibnamefont
  {Kn\"{u}lle}}, \bibinfo {author} {\bibfnamefont {P.}~\bibnamefont {Fischer}},
  \bibinfo {author} {\bibfnamefont {S.}~\bibnamefont {Parkin}}, \ and\ \bibinfo
  {author} {\bibfnamefont {H.}~\bibnamefont {Ebert}},\ }\href@noop {}
  {\bibfield  {journal} {\bibinfo  {journal} {Journal of Applied Physics}\
  }\textbf {\bibinfo {volume} {73}},\ \bibinfo {pages} {6430} (\bibinfo {year}
  {1993})}\BibitemShut {NoStop}%
\bibitem [{\citenamefont {Bertero}\ and\ \citenamefont
  {Sinclair}(1994)}]{berteroJMMM94}%
  \BibitemOpen
  \bibfield  {author} {\bibinfo {author} {\bibfnamefont {G.}~\bibnamefont
  {Bertero}}\ and\ \bibinfo {author} {\bibfnamefont {R.}~\bibnamefont
  {Sinclair}},\ }\href@noop {} {\bibfield  {journal} {\bibinfo  {journal}
  {Journal of Magnetism and Magnetic Materials}\ }\textbf {\bibinfo {volume}
  {134}},\ \bibinfo {pages} {173 } (\bibinfo {year} {1994})}\BibitemShut
  {NoStop}%
\bibitem [{\citenamefont {Sakurai}\ and\ \citenamefont
  {Shinjo}(1993)}]{shinjoJMMM93}%
  \BibitemOpen
  \bibfield  {author} {\bibinfo {author} {\bibfnamefont {M.}~\bibnamefont
  {Sakurai}}\ and\ \bibinfo {author} {\bibfnamefont {T.}~\bibnamefont
  {Shinjo}},\ }\href@noop {} {\bibfield  {journal} {\bibinfo  {journal}
  {Journal of Magnetism and Magnetic Materials}\ }\textbf {\bibinfo {volume}
  {128}},\ \bibinfo {pages} {237} (\bibinfo {year} {1993})}\BibitemShut
  {NoStop}%
\bibitem [{\citenamefont {Ghosh}\ \emph {et~al.}(2012)\citenamefont {Ghosh},
  \citenamefont {Auffret}, \citenamefont {Ebels},\ and\ \citenamefont
  {Bailey}}]{ghoshPRL12}%
  \BibitemOpen
  \bibfield  {author} {\bibinfo {author} {\bibfnamefont {A.}~\bibnamefont
  {Ghosh}}, \bibinfo {author} {\bibfnamefont {S.}~\bibnamefont {Auffret}},
  \bibinfo {author} {\bibfnamefont {U.}~\bibnamefont {Ebels}}, \ and\ \bibinfo
  {author} {\bibfnamefont {W.}~\bibnamefont {Bailey}},\ }\href@noop {}
  {\bibfield  {journal} {\bibinfo  {journal} {Phys. Rev. Lett.}\ }\textbf
  {\bibinfo {volume} {109}},\ \bibinfo {pages} {032000} (\bibinfo {year}
  {2012})}\BibitemShut {NoStop}%
\bibitem [{\citenamefont {Heinrich}\ \emph {et~al.}(2003)\citenamefont
  {Heinrich}, \citenamefont {Tserkovnyak}, \citenamefont {Woltersdorf},
  \citenamefont {Brataas}, \citenamefont {Urban},\ and\ \citenamefont
  {Bauer}}]{heinrich-2003}%
  \BibitemOpen
  \bibfield  {author} {\bibinfo {author} {\bibfnamefont {B.}~\bibnamefont
  {Heinrich}}, \bibinfo {author} {\bibfnamefont {Y.}~\bibnamefont
  {Tserkovnyak}}, \bibinfo {author} {\bibfnamefont {G.}~\bibnamefont
  {Woltersdorf}}, \bibinfo {author} {\bibfnamefont {A.}~\bibnamefont
  {Brataas}}, \bibinfo {author} {\bibfnamefont {R.}~\bibnamefont {Urban}}, \
  and\ \bibinfo {author} {\bibfnamefont {G.}~\bibnamefont {Bauer}},\ }\href
  {http://dx.doi.org/10.1103/PhysRevLett.90.187601} {\bibfield  {journal}
  {\bibinfo  {journal} {Physical Review Letters}\ }\textbf {\bibinfo {volume}
  {90}},\ \bibinfo {pages} {187601 } (\bibinfo {year} {2003})}\BibitemShut
  {NoStop}%
\bibitem [{\citenamefont {Brataas}\ \emph {et~al.}(2000)\citenamefont
  {Brataas}, \citenamefont {Nazarov},\ and\ \citenamefont
  {Bauer}}]{brataasPRL00}%
  \BibitemOpen
  \bibfield  {author} {\bibinfo {author} {\bibfnamefont {A.}~\bibnamefont
  {Brataas}}, \bibinfo {author} {\bibfnamefont {Y.~V.}\ \bibnamefont
  {Nazarov}}, \ and\ \bibinfo {author} {\bibfnamefont {G.~E.~W.}\ \bibnamefont
  {Bauer}},\ }\href@noop {} {\bibfield  {journal} {\bibinfo  {journal} {Phys.
  Rev. Lett.}\ }\textbf {\bibinfo {volume} {84}},\ \bibinfo {pages} {2481}
  (\bibinfo {year} {2000})}\BibitemShut {NoStop}%
\bibitem [{\citenamefont {Zwierzycki}\ \emph {et~al.}(2005)\citenamefont
  {Zwierzycki}, \citenamefont {Tserkovnyak}, \citenamefont {Kelly},
  \citenamefont {Brataas},\ and\ \citenamefont {Bauer}}]{bauerPRB05}%
  \BibitemOpen
  \bibfield  {author} {\bibinfo {author} {\bibfnamefont {M.}~\bibnamefont
  {Zwierzycki}}, \bibinfo {author} {\bibfnamefont {Y.}~\bibnamefont
  {Tserkovnyak}}, \bibinfo {author} {\bibfnamefont {P.~J.}\ \bibnamefont
  {Kelly}}, \bibinfo {author} {\bibfnamefont {A.}~\bibnamefont {Brataas}}, \
  and\ \bibinfo {author} {\bibfnamefont {G.~E.~W.}\ \bibnamefont {Bauer}},\
  }\href {\doibase 10.1103/PhysRevB.71.064420} {\bibfield  {journal} {\bibinfo
  {journal} {Phys. Rev. B}\ }\textbf {\bibinfo {volume} {71}},\ \bibinfo
  {pages} {064420} (\bibinfo {year} {2005})}\BibitemShut {NoStop}%
\bibitem [{\citenamefont {Costa}\ \emph {et~al.}(2006)\citenamefont {Costa},
  \citenamefont {Muniz},\ and\ \citenamefont {Mills}}]{costaPRB06}%
  \BibitemOpen
  \bibfield  {author} {\bibinfo {author} {\bibfnamefont {A.}~\bibnamefont
  {Costa}}, \bibinfo {author} {\bibfnamefont {R.}~\bibnamefont {Muniz}}, \ and\
  \bibinfo {author} {\bibfnamefont {D.}~\bibnamefont {Mills}},\ }\href@noop {}
  {\bibfield  {journal} {\bibinfo  {journal} {Phys. Rev. B}\ }\textbf {\bibinfo
  {volume} {73}},\ \bibinfo {pages} {054426} (\bibinfo {year}
  {2006})}\BibitemShut {NoStop}%
\bibitem [{l-b(1990{\natexlab{c}})}]{l-b-eband-book}%
  \BibitemOpen
  \href@noop {} {\emph {\bibinfo {title} {Landolt-Bornstein Tables}}}\
  (\bibinfo  {publisher} {Springer Verlag},\ \bibinfo {year}
  {1990})\BibitemShut {NoStop}%
\end{thebibliography}
\end{document}